\documentclass[aps,prx,reprint,superscriptaddress,footnote, twocolumn,notitlepage]{revtex4-1}
\usepackage{amsmath, amsthm, amssymb,amsfonts,mathbbol,amstext}
\usepackage{graphicx}
\usepackage{dcolumn}
\usepackage{bm}
\usepackage{bbm}
\usepackage[colorlinks,linkcolor=blue,urlcolor=blue, citecolor=blue]{hyperref}
\usepackage{mathtools}
\usepackage{color}
\usepackage{multirow}
\usepackage{diagbox}
\usepackage{float}
\usepackage{xcolor}
\usepackage{listings}

\usepackage{changes}

\def\1{\mathbf{1}}
\def\0{\mathbf{0}}

\newcommand*{\coloneqq}{\mathrel{\vcenter{\baselineskip0.5ex \lineskiplimit0pt \hbox{\scriptsize.}\hbox{\scriptsize.}}} =}

\newcommand{\processnext}[1]{%
	\ifx\listfinish#1\empty\else\listact{#1}\expandafter\processnext\fi}

\newcommand{\ignore}[1]{}
\newcommand{\nobibentry}[1]{{\let\nocite\ignore\bibentry{#1}}}

\newcommand{\bibfnamefont}[1]{#1}
\newcommand{\bibnamefont}[1]{#1}

\let\oldsqrt\sqrt
\def\sqrt{\mathpalette\DHLhksqrt}
\def\DHLhksqrt#1#2{\setbox0=\hbox{$#1\oldsqrt{#2\,}$}\dimen0=\ht0
	\advance\dimen0-0.2\ht0
	\setbox2=\hbox{\vrule height\ht0 depth -\dimen0}%
	{\box0\lower0.4pt\box2}}

\begin{document}
	\title{Optimal limits of continuously monitored thermometers and their Hamiltonian structure}
	\author{Mohammad Mehboudi}
    \email{mohammad.mehboudi@tuwien.ac.at}
	\address{Technische Universität Wien, 1020 Vienna, Austria}
    \author{Florian Meier}
    \email{florian.meier@tuwien.ac.at}
	\address{Technische Universität Wien, 1020 Vienna, Austria}
    \author{Marcus Huber}
    \email{marcus.huber@tuwien.ac.at}
	\address{Technische Universität Wien, 1020 Vienna, Austria}
 \address{Institute for Quantum Optics and Quantum Information (IQOQI), Austrian Academy of Sciences, Boltzmanngasse 3, 1090 Vienna, Austria}
    \author{Harry J. D. Miller}
    \email{harry.miller@manchester.ac.uk}
	\address{Department of Physics and Astronomy, University of Manchester, Oxford Road, Manchester M13 9PL, United Kingdom}
	\begin{abstract}
     We investigate the fundamental and practical precision limits of thermometry in bosonic and fermionic environments by coupling an $N$-level probe to them and continuously monitoring it. Our findings show that the ultimate precision limit, quantified by the Fisher information, scales linearly with $N$, offering an exponential improvement over equilibrium thermometry, where the scaling is only $\log^2 N$.
    For a fixed Hamiltonian structure, we develop a maximum likelihood estimation strategy that maps the observed continuously monitored trajectories of the probe into temperature estimates with minimal error. By optimizing over all possible Hamiltonian structures, we discover that the optimal configuration is an effective two-level system, with both levels exhibiting degeneracy that increases with $N$---a stark contrast to equilibrium thermometry, where the ground state remains non-degenerate.
    Our results have practical implications. First, continuous monitoring is experimentally feasible on several platforms and accounts for the preparation time of the probe, which is often overlooked in other approaches such as prepare-and-reset. 
    Second, the linear scaling is robust against deviations from the effective two-level structure of the optimal Hamiltonian.
    Additionally, this robustness extends to cases of initial ignorance about the temperature. Thus, in global estimation problems, the linear scaling remains intact even without adaptive strategies.
     
	\end{abstract}
	\maketitle
{\it Introduction.---} High precision temperature measurements are crucial for the development of nanoelectronic \cite{levitin2022cooling,sarsby2020500} and cold-atom technologies \cite{bloch2005ultracold,chen2020emergence}, with accurate low-temperature sensing now fundamental for  experimental tests in condensed matter \cite{nayak2008non,hasan2010colloquium}, many-body thermalisation \cite{langen2015ultracold,langen2013local} and the realisation of small-scale thermodynamic devices \cite{myers2022quantum}. Recent thermometry techniques have been proposed across a wide range of different systems, such as solid-state impurities \cite{mihailescu2023thermometry} optomechanical setups \cite{purdy2015optomechanical}, quantum gases \cite{PhysRevLett.122.030403,oghittu2022quantum,khan2022subnanokelvin,PhysRevLett.128.040502}, dephased impurities \cite{mitchison2020situ} and topological spinless fermions \cite{srivastava2023topological}. However, there remains an ongoing search to find methods for probing the temperatures of many-body systems that are simultaneously practical, accurate and resource efficient \cite{Mehboudi_2019}. This ultimately entails designing a small probe that can interact with the sample and encode information about its temperature in an optimal manner, with estimation schemes established through either local \cite{PhysRevLett.114.220405,de2017universal,de2018quantum,PhysRevB.98.045101,Mehboudi_2019,glatthard2023energy,Potts2019fundamentallimits,PhysRevResearch.2.033394} or global methods \cite{rubio2021global,glatthard2022optimal,alves2022bayesian,mehboudi2022fundamental,PhysRevA.104.052214}. One common approach is \textit{equilibrium thermometry} \cite{stace2010quantum,mok2021optimal,walker2003quantum,carlos2015thermometry,plodzien2018few,nguyen2018all}, where temperature estimates are obtained via energy measurements of a thermalised probe. It has been shown that the optimal energy structure of such an equilibrium thermometer is an effective $2$-level system with a single ground state and a maximally degenerate ($N-1$ fold) exited state at an optimal gap $\epsilon_{\rm eq}^*$ \cite{PhysRevLett.114.220405}. Subsequently, several studies have addressed how to simulate such energy structure by using physically realisable many-body models~\cite{PhysRevA.92.052112,plodzien2018few,PhysRevA.90.022111,Mehboudi2015,SaladoMeja2021,Mok2021} and even getting the right scaling and asymptotic behaviour with $2$-body interactions~\cite{arxiv.2211.01934,arxiv.2311.14524}. On the other hand, the degenerate energy structure also leads to long equilibration timescales for the probe \cite{ivander2023hyperacceleration}; if one factors in time as a resource in metrology \cite{dooley2016quantum,hayes2018making}, equilibrium thermometry may not be the preferable approach \cite{anto2023bypassing}. This instead motivates the use of finite-time probes that encode information about temperature via non-equilibrium dynamics \cite{brunelli2011qubit,cavina2018bridging,sekatski2022optimal,luiz2022machine,rodriguez2024strongly,2407.21618,PhysRevA.109.023309,PhysRevLett.123.180602}. However, in these scenarios the question of optimality is not so clear, particularly due to the need to use adaptive estimation schemes, repeatable state preparations alongside measurements at precise times.   

The aim of this paper is to explore optimal thermometry in an alternative approach based on \textit{continuous monitoring} \cite{gammelmark2014fisher,kiilerich2014estimation,rossi2020noisy,ALBARELLI2024129260}. This is a practical and convenient method that involves monitoring the probe populations across time while out of equilibrium \cite{boeyens2023probe}. Since the transition rates governing the dynamics of the probe depend on temperature, one can apply a maximum-likelihood estimation scheme directly to the population trajectories~\cite{smiga2023stochastic}. This clearly has a resource advantage over other approaches that reset or discard the probe after each measurement, and instead makes full use of the information contained in the dynamical history of a \textit{single} probe. With recent advancements in continuous monitoring methods such as photo-detection, homodyne/heterodyne detection \cite{zhang2017quantum,vijay2012stabilizing} and other quantum non-demolition
measurements \cite{Kong2020}, this approach to thermometry is experimentally realisable. 

Given the feasibility of the continuous monitoring approach along with its advantages in terms of resources, one remaining goal is to find the best energy structure for the probe. Using a combination of both analytic and numerical methods, we demonstrate that the optimal structure mirrors that of an equilibrium thermometer, i.e., an effective $2$-level system with a degenerate ground state and a degenerate excited state. Investigating both fermionic and bosonic samples, we show that the precision of an $N$-level probe can achieve linear scaling with the number of levels, thereby providing an exponential improvement to equilibrium thermometry methods such as \cite{PhysRevLett.114.220405}. Furthermore, this optimal structure and its performance bounds apply universally across all temperature ranges, thus bypassing the need for adaptive measurement scheme such as \cite{mehboudi2022fundamental}.  Overall, our results solidify the significant advantages offered by continuously monitored thermometers.

\begin{figure}
    \centering
    \includegraphics[width=0.9\linewidth]{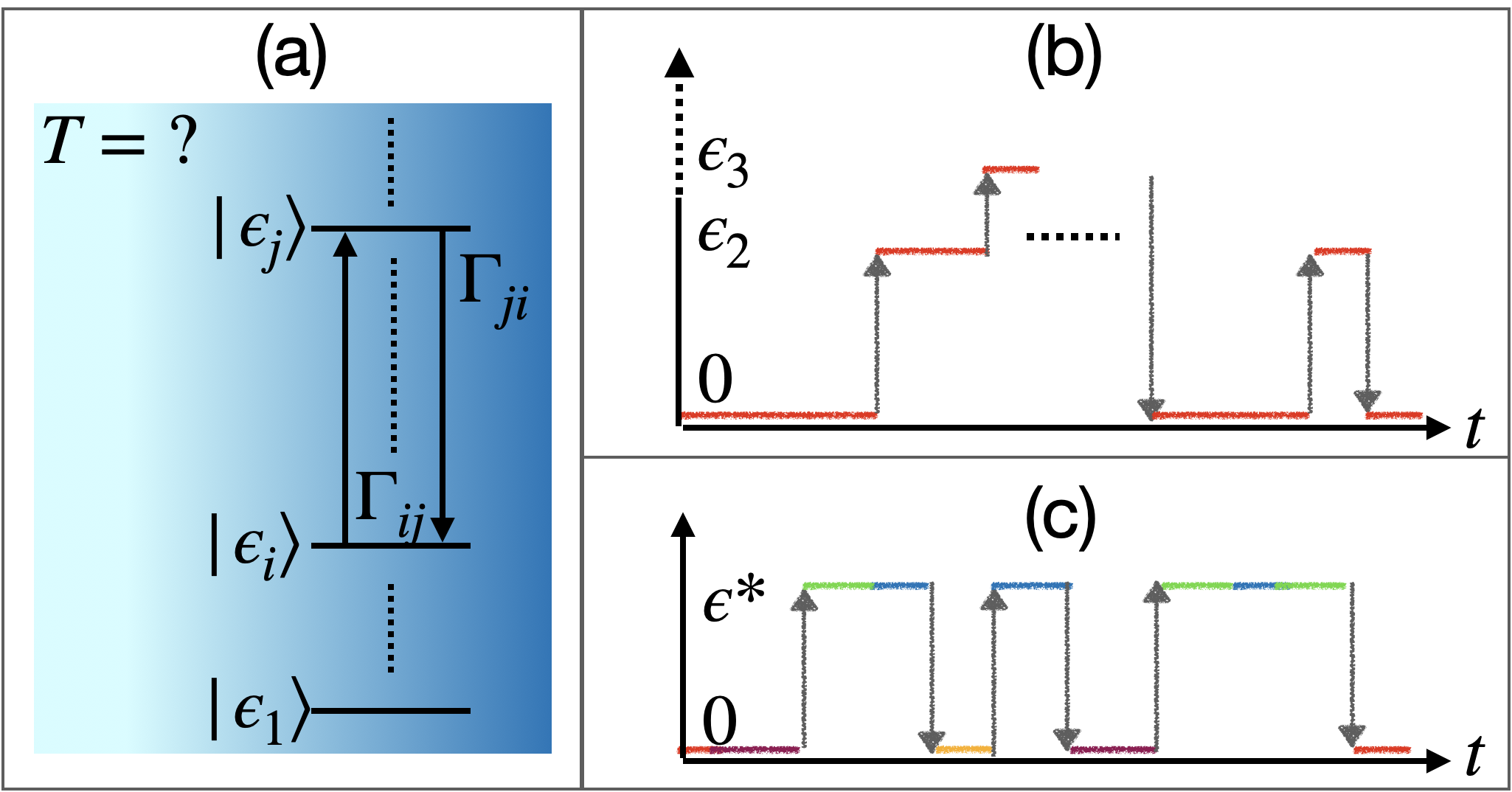}
    \caption{(a) An $N$-level probe is in contact with a thermal bath at temperature $T$. The transition rates between different energy levels, $\Gamma_{ij}$, are temperature dependent. (b) The trajectory that the $N$-level system takes contains information about the temperature. The precision with which the trajectory can estimate the temperature highly depends on the energy structure. (c) The optimal energy structure---as we find in this work---is effectively 2-level, with ground state and exited state degeneracy that both increase linearly with $N$, and at some optimal gap $\epsilon^*$. In the figure we have color coded the degenerate gaps to tell them apart visually.}
    \label{fig:enter-label}
\end{figure}

\textit{Setup.}-- We will begin by modelling a probe as a finite-dimensional system with $N$ energy levels, which is in weak contact with a thermal sample at inverse temperature $\beta=1/k_B T$. It is assumed that the population dynamics of the probe follow a Markovian rate equation of typical form
\begin{align}\label{eq:rate}
    \dot{\mathbf{p}}(t) = \mathbf{p}(t)\Gamma,
\end{align}
with $\mathbf{p}(t)=[p_1(t),...,p_N(t)]$ being the vector of populations of the energy levels $\{\epsilon_1,\epsilon_2,...\epsilon_N\}$, and $\Gamma$ being the transition matrix. The matrix elements $\Gamma_{ij}$ quantify the rate of jumping from state $i$ into the state $j$, with conservation of probability ensured by $\Gamma_{ii} = -\sum_{j\neq i} \Gamma_{ij}$. The rates are assumed to satisfy detailed balance, $\Gamma_{ij}= e^{-\beta(\epsilon_j - \epsilon_i)}\Gamma_{ji}$, so that the thermal distribution $\mathbf{p}^{eq}=\{e^{-\beta \epsilon_1}/Z,..., e^{-\beta \epsilon_N}/Z\}$ is stationary, $\mathbf{p}^{eq}\Gamma =0 $, and we further assume this is unique.

In order to use the probe as a thermometer for estimating temperature $T$, we will follow the scheme introduced in \cite{boeyens2023probe} and continuously monitor its energy over time. This is a jump unravelling process~\cite{PRXQuantum.5.020201}, and the schematics of this setup is shown in Figure~\ref{fig:enter-label}. The data output of the continuous monitoring over some time window $[0,\tau]$ is represented by a stochastic trajectory of a discrete-time process $\mathbf{X}_\tau=\{(n_0,t_0=0),(n_1,t_1),...,(n_m,t_m=\tau)\}$, where each $(n_k,t_k)$ denotes an observation of the $n_k$'th energy level $\epsilon_{n_k}$ at an intermediate time $ 0\leq t_k \leq \tau$, where $t_k = k \tau/m$. The probability of this trajectory is of the form $P(\mathbf{X}_\tau)=p(n_0)p(n_1|n_0)...p(n_{m}|n_{m-1})$ with each transition probability determined by the rates in the master equation~\eqref{eq:rate}. Since the trajectory probability is dependent upon the temperature of the sample, $P(\mathbf{X}_\tau)=P(\mathbf{X}_\tau|T)$, we can use the data to construct an estimator $\hat{T}$. For large $\tau$ a suitable estimator becomes unbiased and its maximum precision is fixed by the Cramer-Rao bound, which bounds the mean-squared error $\text{Var} \ (\hat{T})\geq 1/F(T)$ in terms of the Fisher information (FI), defined $F(T):=\langle [\partial_T \text{ln} P(\mathbf{X}_\tau|T)]^2 \rangle$. As the system obeys an order-$1$ Markov process with unique fixed point $\mathbf{p}^{eq}$, we can use a result from \cite{smiga2023stochastic} to express the Fisher information directly in terms of the master equation rates and thermal populations $p^{eq}_i=e^{-\beta \epsilon_i}/Z$,
\begin{align}\label{eq:FI}
    F(T)=\tau \beta^4 \sum_i p^{eq}_i \sum_{j\neq i}\frac{|\partial_{\beta} \Gamma_{ij}|^2}{\Gamma_{ij}}
\end{align}
Our main goal will be to maximise~\eqref{eq:FI} by choosing an optimal energy structure of the probe, which determines the functional behaviour of the rates $\Gamma_{ij}$. A key figure of merit will be its scaling with respect to the number of levels, which is an important cost in terms of the dimensionality of the system. For example, a probe consisting of $n$ 2-level spin states is related to the number of levels as $N=2^n$. One should bear in mind that the optimal scaling of $F(T)$ will at most be linear in $N$ due to the bound
\begin{align}\label{eq:ultimate}
     F(T) \leq \tau \beta^4 (N-1) \max_{\{i,j\}} \frac{|\partial_{\theta} \Gamma_{ij}|^2}{\Gamma_{ij}}.
\end{align}
While this bound is not necessarily tight in a continuously monitored setup, one can saturate it by using a measure-and-reset strategy following a result from~\cite{sekatski2022optimal}.
That is, if we were allowed to reset the system after each jump, we would choose an effective 2-level Hamiltonian structure, with a ground state and an $N-1$ fold degenerate exited state at a gap $\epsilon_j$ chosen such that the fraction $\frac{|\partial_{\theta} \Gamma_{0j}|^2}{\Gamma_{0j}}$ is maximised independent of $N$. After each measurement, one would reset the system to the ground state, which comes with an unwanted resource cost. Fortunately we now demonstrate that the linear scaling suggested by~\eqref{eq:ultimate} is in fact achievable in the continuous approach, avoiding this need to reset. To understand why this is indeed the case, note that $\dots$

\begin{figure}
    \centering
    \includegraphics[width=0.95\linewidth]{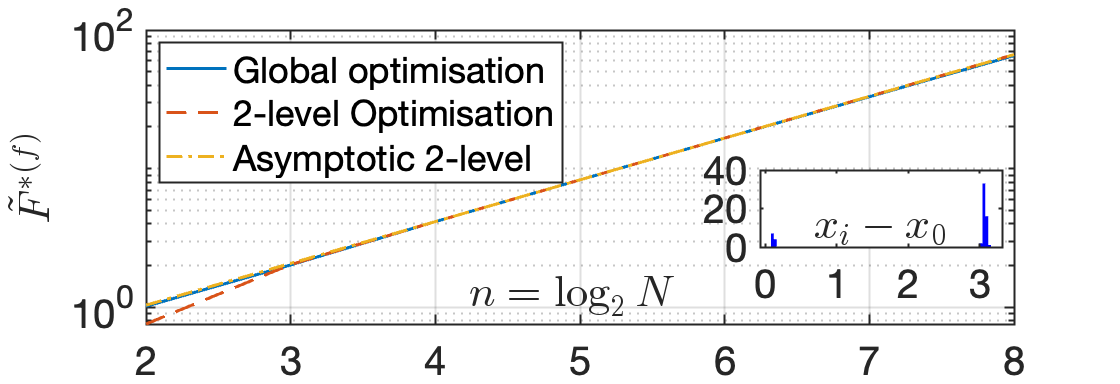}
    \caption{Numerical results for the optimal dimensionless FI rate~\eqref{eq:FIrate}. We compare the exact solution obtained via global optimisation with the optimal $2$-level ansatz~\eqref{eq:2level} (optimised over $N_0$ and $x$). The analytic value~\eqref{eq:fermion_asympt} obtained in the asymptotic $N\gg 1$ limit is also plotted and shows good agreement with the exact solution for $n\geq 3$. \textbf{Inset}: This plot depicts a histogram of the energy levels found from a global optimisation for $n=6$. 
    Our numerics suggest that an effective $2$-level structure is optimal. We have thus also considered an effective $2$-level ansatz which can be optimised more efficiently due to fewer parameters. The FI rate of the $2$-level ansatz is shown by the red curve in the primary figure which also as one can see agrees very well with the global optimum for $n\geq 3$.
    } \label{fig:global_optimisation_fermionic}
\end{figure}

\begin{figure*}
    \centering
         \hspace{-.7cm}
         \includegraphics[width=0.42\linewidth]{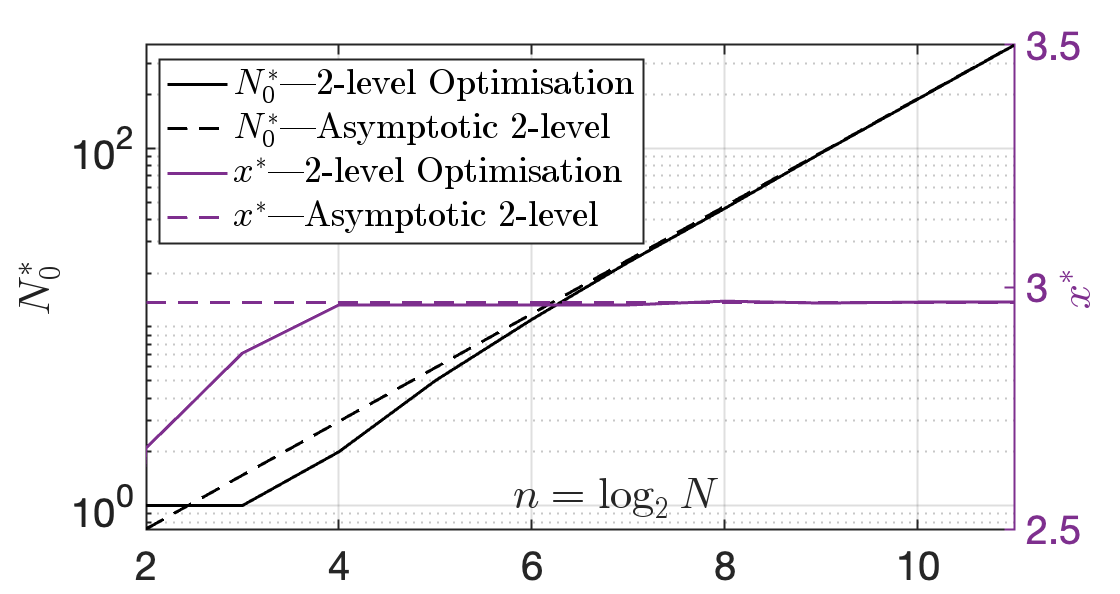}
    \includegraphics[width=0.42\linewidth]{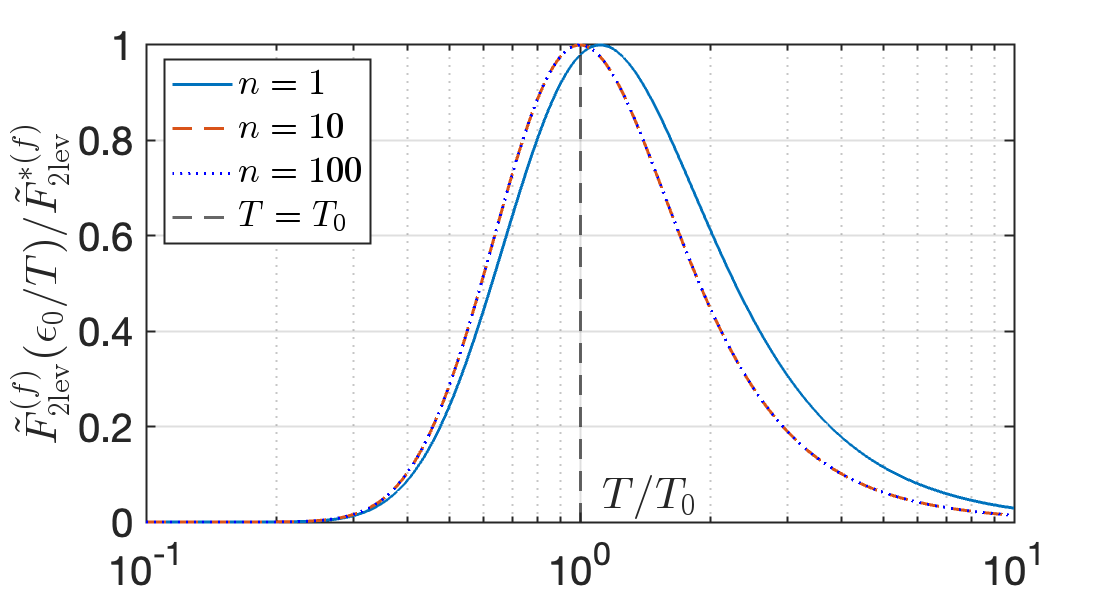}
    \caption{\textbf{Left:} Plot of the optimal free parameters for the $2$-level ansatz~\eqref{eq:2level} for a fermionic probe as a function of $n=\log_2 N$, with optimal ground state degeneracy $N_0^*$ (black) and optimal gap $x^*$ (red). The dashed lines give the values for the asymptotic optimal solution~\eqref{eq:fermion_asympt}. \textbf{Right:} Plot of FI rate normalised by its maximum value, for different values of $n$. Here the gap is tuned to $\epsilon_0 = x^* T_0$, such that the thermometer is optimal for some assumed temperature $T_0$.}
    \label{fig:2levans_ferm}
\end{figure*}

\textit{Fermionic sample.}---We first look at the estimation of a fermionic sample with a spectral density taken in the wide-band limit. The transition rates of the probe are proportional to the Fermi occupation $n_F(\omega_{ji})$ of each excitation $\omega_{ji} \equiv \epsilon_j - \epsilon_i$, namely
\begin{align}
\label{eq:Gamma_ij_Fermionic}
\Gamma_{ij} =\gamma n_F(\omega_{ji})=\gamma [e^{\beta\omega_{ji}} + 1]^{-1}
\end{align}
with $\gamma$ the coupling strength. It follows from~\eqref{eq:FI} that the FI is
\begin{align}\label{eq:ferm_FI_beta}
    F(T) = \gamma\tau\beta^4 \sum_i p^{eq}_i \sum_{j\neq i}\omega_{ji}^2\frac{e^{2\beta\omega_{ji}}}{[1+e^{\beta\omega_{ji}}]^3}=\gamma\tau\beta^2 \tilde{F}^{(f)}
\end{align}
where we have identified the dimensionless FI rate, 
\begin{align}\label{eq:FIrate}
    {\tilde F}^{(f)} \coloneqq  \sum_i p^{eq}_i \sum_{j\neq i}x_{ji}^2\frac{e^{2 x_{ji}}}{[1+e^{x_{ji}}]^3},
\end{align}
and set $x_{\ij} = \beta \omega_{ij}$. The goal is now to optimise ${\tilde F}^{(f)}$ over the set of  $\{x_i=\beta\epsilon_i\}_{i=0}^{N-1}$.
As a non-linear problem, we first turn to numerics and run an optimisation in matlab~\footnote{All matlab codes are publicly available on \href{https://github.com/Mehboudi/Thermometry-with-Continious-Monitoring-using-N-level-probes.git}{this GitHub repository}.}. For $N=2^n$ with $n\in\{2,\dots,8\}$ we observe two things: (1) The optimal value ${\tilde F}^{*(f)} = \max_{\{x_{i}\}} {\tilde F}^{(f)}$ grows linearly with $N$. (2) For $n\geq 3$, an effective 2-level degenerate structure is able to reach the global optimum with small numerical error---see Figure~\ref{fig:global_optimisation_fermionic}. We therefore adopt an \textit{ansatz} consisting of an $N_0$ ground state levels and $N-N_0$ excited state levels, with a single dimensionless gap $x=\beta\omega$. In this case the FI rate reads
\begin{align}\label{eq:2level}
    {\tilde F}_{2\rm lev}^{(f)} = \frac{N_0(N-N_0)}{N_0 + (N-N_0)e^{-x}}\left(\frac{x^2e^{2 x}}{[1+e^{x}]^3} + \frac{x^2e^{-3 x}}{[1+e^{-x}]^3}\right),
\end{align}
To optimise further, there are two remaining free parameters  $N_0$ and $x$. Figure~\ref{fig:2levans_ferm} shows that
$x$ is tending to a constant for large $n$, while the degeneracy appears to exponentially increase with $n$. To see this analytically we look at the $N\gg 1$ regime and treat $N_0$ as a continuous number, so that it can be expressed as a fraction $N_0 = C N$. Straightforwardly one can show that, for a fixed $x$, the optimal solution of $C$ depends on the gap and reads $C(x)=1/(1+e^{x/2})$. Substituting back in the FI rate, we get
\begin{align}\label{eq:fermion_asympt}
    {\tilde F}_{2\rm lev}^{(f)} & = N 
    x^2 \frac{C(x)(1-C(x))}{C(x) + (1-C(x))e^{-x}} \, \nonumber\\
    & \times {\left(\frac{{\mathrm{e}}^{2\,x} }{{{\left({\mathrm{e}}^x +1\right)}}^3 }+\frac{{\mathrm{e}}^{-3\,x} }{{{\left({\mathrm{e}}^{-x} +1\right)}}^3 }\right)}\,
    = N f(x)
\end{align}
Interestingly, the function $f(x)$ has a unique maximum which can be found numerically. Therefore, for large enough $N$, we have the optimal structure and its corresponding Fisher information rate as ${\tilde F}_{2\rm lev}^{*(f)}  \approx 0.2596 N$, $x^*  \approx 2.9682$ and $N_0^*  \approx \max\{{\rm floor}(0.1848 ~ N),~1\}$. As a comparison, note that the bound in Eq.~\eqref{eq:ultimate} that requires resetting the probe after each measurement, gives $x_{\rm reset}^* \approx 2.5331$, and $F^{*(f)}_{\rm reset} \approx 0.4052 N$. Finally, a key observation is that the width of the Fisher information for the optimal continuously monitored thermometer does not shrink with $n=\log_2 N$, it actually remains constant, as can be seen from its expression above and also Figure~\ref{fig:2levans_ferm}. Thus avoiding the fate of the optimal equilibrium thermometer~\cite{PhysRevLett.114.220405} when our initial knowledge of the temperature is limited---that is global thermometry. While using adaptive schemes as proposed in~\cite{PhysRevLett.128.130502} may improve the precision up to a coefficient, the scaling is not compromised if one does not use them. In Appendix D, we furthermore show that even if the degenerate energy levels are not perfectly prepared, the linear scaling survives, adding an additional layer of robustness and practicality into the optimal scaling.

\begin{figure*}
    \centering
       \includegraphics[width=0.42\linewidth]{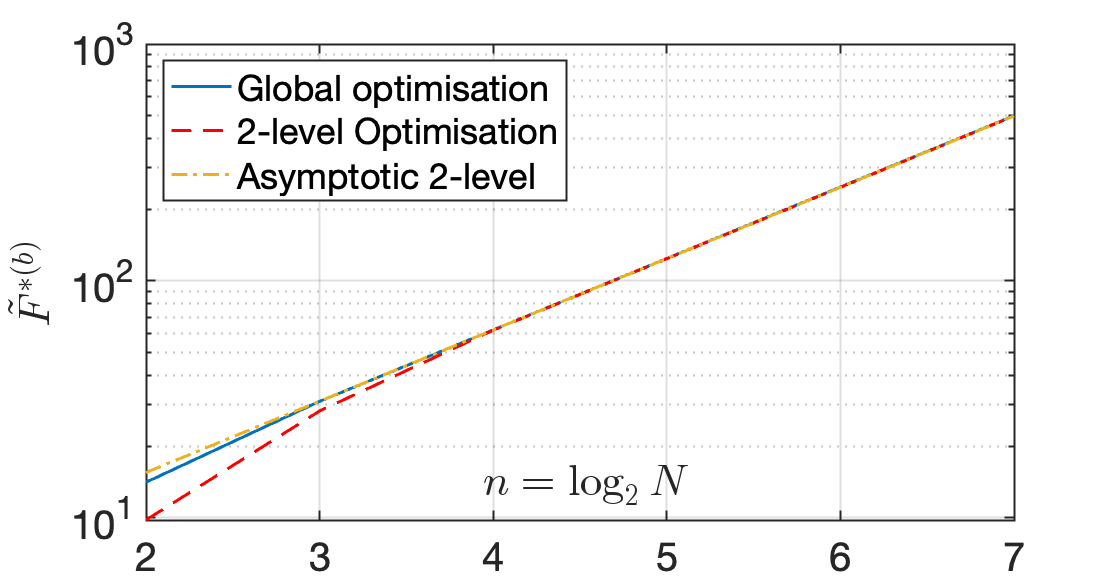}
       \includegraphics[width=0.42\linewidth]{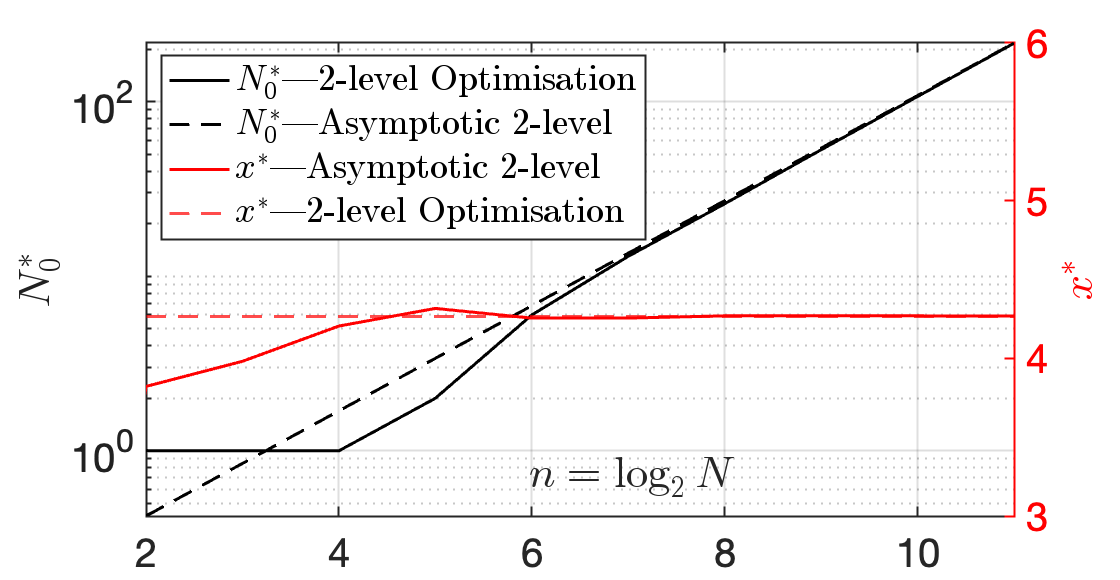}

    \caption{\textbf{Left}: the optimal FI rate~\eqref{eq:FIrateboson} for a bosonic case with ohmicity $s=2$, using a global optimisation (solid) and the 2-level ansatz~\eqref{eq:2levelboson} (red-dashed). The asymptotic values are taken from Table~\ref{table:bos_opt}. \textbf{Right:} The optimal energy gap $x^*$ (red) for the 2-level ansatz and its asymptotic prediction from Table~\ref{table:bos_opt}  and the ground state degeneracy of the optimal structure for the  2-level ansatz (black) and its asymptotic behaviour from Table~\ref{table:bos_opt}.}
    \label{fig:bosonic_2level}
\end{figure*}

\textit{Bosonic sample.}---We now investigate the performance of the probe when coupled to a bosonic sample. In this case the transition rates are given by $\Gamma_{ij} = \kappa(|\omega_{ji}|) |n_B(\omega_{ji})|$ where  $n_B(\omega) = [e^{\beta\omega} - 1]^{-1}$ is the Bose-Einstein distribution and $\kappa(\omega)$ represents the spectral density. Here we restrict attention to super Ohmic spectral densities of the form $\kappa(\omega) = \gamma\omega^{s}$ with $s >  1$. 
 
In analogy with~\eqref{eq:FIrate}, the figure of merit for the probe performance is the FI rate
\begin{align}\label{eq:FIrateboson}
    {\tilde F}^{(b)}:=\frac{F(T)}{\gamma \tau T^{2+s}}= \sum_i p^{eq}_i \sum_{j\neq i} \frac{x_{ji}^{2+s}e^{2x_{ji}}}{|e^{x_{ji}}-1|^3},
\end{align} 
where we combined~\eqref{eq:FI} with the bosonic decay rates. Note that in this case, the non-flat spectral density means that we must divide by an additional factor $T^s$ to preserve scale invariance \cite{rubio2022quantum}.  

\begin{table}[h]
    \centering
    \begin{tabular}{cccc}
    \hline\hline
        SD & $x^{*}$ & $C^{*}$ & $b(x^*)$\\
        \hline\hline
        $s=1^+$ & 3.0880 & 0.1760 & 1.0508\\
        \hline
        $s=1.5$ & 3.7195 & 0.1347 & 1.9403\\
        \hline
        $s=2$ & 4.2681 & 0.1058 & 3.8782 \\
        \hline
        $s=3$ & 5.2706 & 0.0669 & 18.4880\\
        \hline
    \end{tabular}
    \caption{The optimal values of the gap and the corresponding ground state degeneracy fraction $C^*$ and the FI coefficient $b(x^*)$ defined in~\eqref{eq:FIrateboson2} for various choices of spectral density (SD). The case with $s=1^+$ is calculated for $s=1+{\epsilon}$ and then taken at the limit of $\epsilon\to 0$.}
    \label{table:bos_opt}
\end{table}

Similar to the fermionic case, numerics suggest again that---for $n>3$---a 2-level energy structure globally optimises~\eqref{eq:FIrateboson}. Taking a 2-level ansatz for~\eqref{eq:FIrateboson} gives
\begin{align}\label{eq:2levelboson}
    {\tilde F}_{2\rm lev}^{(b)} = \frac{N_0(N-N_0)}{N_0 + (N-N_0)e^{-x}} \left(\frac{x^{2+s}e^{2x}}{|e^{x}-1|^3 } + \frac{x^{2+s}e^{-3x}}{|e^{-x}-1|^3} \right),
\end{align}
We numerically optimise this 2-level structure to find the optimal $N_0$ and $x$ as depicted in Fig.~\ref{fig:bosonic_2level}, and see similar qualitative behaviour as in the fermionic case. Looking at the asymptotic regime $N\gg 1$ as we did with~\eqref{eq:fermion_asympt}, the FI rate is
\begin{align}\label{eq:FIrateboson2}
     {\tilde F}_{2\rm lev}^{(b)} & = N \frac{C(x)(1-C(x))}{C(x) + (1-C(x))e^{-x}}\nonumber\\
     &  \ \ \ \ \times \left(\frac{x^{2+s}e^{2x}}{|e^{x}-1|^3 } + \frac{x^{2+s}e^{-3x}}{|e^{-x}-1|^3} \right) 
    = N b(x)
\end{align}
 The ohmicity parameter is determinant on the optimal gap, the ground state degeneracy, and the corresponding FI. These are summarised in the Table~\ref{table:bos_opt} for various ohmicity parameters. 
Again, if we compare with the reset-and-measure setting bound in Eq.~\ref{eq:ultimate}, we would get $F_{\rm reset}^{*(b)} = \{1.6786, 2.7144, 4.9953, 21.5120\}N$ and $x_{\rm reset}^{*} = \{2.7144, 3.0430, 3.7240, 4.8890\}$ respectively for $s=\{1^+, 1.5, 2, 3\}$.

\textit{Discussion}-- We have demonstrated that the optimal energy structure of a continuously monitored thermometer, for both fermionic and bosonic samples, is an effective two-level system that can achieve linear scaling with respect to the number of levels. To complete the picture, we also derive the corresponding maximum likelihood estimator for this setup in Appendix~\ref{app:A}, which allows saturation of the Cramer-Rao bound used implicitly throughout our analysis. There are two distinct factors that highlight the significant advantages offered by the continuous monitoring approach. Firstly, there is a clear exponential improvement over the equilibrium thermometer; in the latter case the FI is known to scale with $(\text{log} N)^2$, as we recap in Appendix~\ref{app:B}. Secondly, we have found that the width of the FI remains constant with increasing probe size. This means that it is possible to achieve a linear scaling using just a single probe without requiring an adaptive feedback approach. Finally, we show in Appendix~\ref{app:C} that the optimal probe performs similarly even when one only has access to the average populations, as quantified in terms of the emperical FI rate \cite{smiga2023stochastic}. This adds another layer of robustness to the results. In summary, taking into account the various resources needed for optimal thermometry, our findings suggest that the continuous monitoring approach performs best. 

It should be emphasised that these results are predicated on the assumption of the Markovian rate equation~\eqref{eq:rate}. An interesting future direction would be to explore the role of non-Markovian effects on the performance of the optimal monitored thermometer. We have also neglected to study thermometry of sub-Ohmic samples since such systems are not well described by~\eqref{eq:rate}. 
A promising subject for future studies is thus to consider non-Markovian dynamics beyond the rate-equation. 

{\it Acknowledgements.---}The authors acknowledge TU Wien Bibliothek for financial support
through its Open Access Funding Programme. This research was funded in part by the Austrian Science Fund (FWF) [10.55776/I6047]. For open access purposes, the author has applied a CC BY public copyright license to any author accepted manuscript version arising from this submission. This research was supported by the European Research Council (Consolidator grant `Cocoquest' 101043705). H. J. D. M. acknowledges funding from a Royal Society Research Fellowship (URF/R1/231394).
This project is co-funded by the European Union (Quantum Flagship project ASPECTS, Grant Agreement No.\ 101080167). 
Views and opinions expressed are however those of the authors only and do not necessarily reflect those of the European Union, REA or UKRI. Neither the European Union nor UKRI can be held responsible for them.

\bibliographystyle{apsrev4-1}
\bibliography{Refs}

\onecolumngrid

\appendix
\vspace{1cm}

\section{The Maximum Likelihood estimator}\label{app:A}
Here, we would like to present the maximum likelihood estimator (MLE) for the temperature in case of the 2-level Hamiltonian structure.
As the transitions between degenerate energy levels do not contribute to the Fisher information, we can coarse grain them in the same bins that we label by $0$ for the $N_0$ ground levels and $1$ for the $N-N_0$ excited levels.
To determine the MLE for the temperature of the bath, we need to look at the probability that observed trajectory has occured.
A stochastic trajectory $\mathbf X_\tau$ is given as in the main text by $\mathbf{X}_\tau=\{(n_0,t_0=0),(n_1,t_1),...,(n_m,t_m=\tau)\}$, which is a list of the states $n_k$ in which the system was at the observation time $t_k$.
In this setting the observation times $t_k=k \tau / m$ are chosen equidistantly in the interval $[0,\tau]$.
The probability of a specific trajectory is then given by a modification of that given in~\cite{boeyens2023probe}---one only should replace $\Gamma_{\rm in} \to (N-N_0)\Gamma_{01}$ and $\Gamma_{\rm out} = N_0\Gamma_{10}$. We have
\begin{align}
\label{eq:X_tau_trajectory_probability}
    \rho(\mathbf X_\tau|T) \propto [(N-N_0)\Gamma_{01}]^k [N_0\Gamma_{10}]^l e^{-(N-N_0)\Gamma_{01} \tau_0} e^{-N_0\Gamma_{10}(\tau - \tau_0)},
\end{align}
where $\tau_0$ is the total time the system spends in the ground state manifold, $k$ is the total jumps from the ground state to the exited state, and $l$ is the total number of jumps from the exited state manifold to the ground state manifold.
Note that this expression~\eqref{eq:X_tau_trajectory_probability} only holds in the case where the observation times are fast, that is $\tau N_0 \Gamma_{10} / m, \tau (N-N_0)\Gamma_{01} / m \ll 1$.

For the MLE estimator of the temperature, we are interested in the temperature $\tilde T_{\rm MLE}$ that maximizes the probability $\rho(\mathbf X_\tau|\tilde T_{\rm MLE})$.
Equivalently, this estimator can be found maximizing the logarithm of the probability, the log-likelyhood function, that is
\begin{align}
    {\tilde T}_{\rm MLE} \coloneqq \arg\max_{T} \log \rho(\nu_\tau|T).
\end{align}
To find $\tilde T_{\rm MLE}$ we can look at the zero solutions of the derivative of the log-likelyhood function, which give us $\tilde T_{\rm MLE}$ in case the solution turns out to be unique.
The derivative in this case can be directly obtained by using Eq.~\eqref{eq:X_tau_trajectory_probability},
\begin{align}
\label{eq:derivtiave_loglikely}
    \partial_T \log \rho(\nu_\tau|T) = k\frac{\partial_T \Gamma_{01}}{\Gamma_{01}} + l\frac{\partial_T \Gamma_{10}}{\Gamma_{10}} -(N-N_0)\tau_0\partial_T \Gamma_{01} - N_0(\tau - \tau_0) \partial_T \Gamma_{10}.
\end{align}
{\it Fermionic bath.---}For
a fermionic bath, let us recall that $\Gamma_{ij} = \gamma / (1+e^{\beta\omega_{ji}})$ is given by the general expression~\eqref{eq:Gamma_ij_Fermionic} from the main text.
For the derivative $\Gamma_{01},$ this results in
\begin{align}
    \frac{\partial_T \Gamma_{01}}{\Gamma_{01}} = \frac{\beta^2 \epsilon}{1+e^{-\beta\epsilon}} = \beta^2\epsilon (1-n_F(x)),
\end{align}
where $\epsilon$ is the energy gap between the ground and excited state, $n_F(x)=(1+e^x)^{-1}$ the Fermionic occupation number and $x=\beta\epsilon$.
The derivative of the rate $\Gamma_{10}$ on the other hand can be found by replacing $\epsilon\rightarrow -\epsilon$ resulting in 
\begin{align}
    \frac{\partial_T\Gamma_{10}}{\Gamma_{10}} = -\beta^2\epsilon n_F(x).
\end{align}
Inserting into Eq.~\eqref{eq:derivtiave_loglikely} and dividing by $\beta^2 \epsilon$, we find
\begin{align}
    0 &= k(1-n_F(x)) - l n_F(x) - (N-N_0)\tau_0 \gamma  n_F(x)(1-n_F(x)) + N_0(\tau-\tau_0) \gamma n_F(x)(1-n_F(x)) \\
    &= k(1-n_F(x)) - l n_F(x) + \xi n_F(x) (1-n_F(x)),
\end{align}
where $\xi = \gamma(N_0\tau - N\tau_0)$.
This quadratic equation can be solved for $n_F(x)$ to yield the solutions
\begin{align}
    \tilde n_{F,\pm} &= \frac{1}{2}\left(1 - \frac{k+l}{\xi} \pm \sqrt{\left(1-\frac{k+l}{\xi}\right)^2 + 4 \frac{k}{\xi}}\right)\\
    &=\frac{1}{2} - \frac{k+l}{2\xi} \pm \frac{1}{2\xi}\sqrt{4k\xi + (k+l-\xi)^2}.
\end{align}
Of the two solutions, only the positive one $\tilde n_F = \tilde n_{F,+}$ is valid (that is $\tilde n_F \in [0,1]$).
From $n_F(x)=(1+e^x)^{-1}$ we can obtain the MLE for the temperature by solving for $T=\epsilon/x,$ which gives
\begin{align}
    {\tilde T}_{\rm MLE}^{(f)} \coloneqq \frac{\epsilon}{\log \frac{1-{\tilde n_F}}{{\tilde n_F}}}.
\end{align}
Note that this expression only gives positive temperatures if $\tilde n_F\leq 1/2$. 

{\it Bosonic bath.---}For
a bosonic bath, we recall the rates are given by $\Gamma_{01}=\kappa(\epsilon) n_B(x)$ and $\Gamma_{10} = \kappa(\epsilon) (n_B(x)+1)$, where $n_B(x) = (e^x-1)^{-1}$ is the Bosonic occupation number.
The derivative terms become
\begin{align}
    \partial_T \Gamma_{10} = \partial_T \Gamma_{01} = \kappa(\epsilon)\partial_T n_B(x),
\end{align}
for both the rates. As a consequence, we can rewrite Eq.~\eqref{eq:derivtiave_loglikely} in the following way and express it as a quadratic equation for $n_B$,
\begin{align}
    0 &= k (n_B(x)+1) + l n_B(x) - \zeta n_B(x)(n_B(x)+1) \\
    \Rightarrow 0 &= n_B(x)^2 + n_B(x)\left(1 - \frac{k+l}{\zeta}\right) - \frac{k}{\zeta}.
\end{align}
Here, $\zeta = \kappa(\epsilon)((N-N_0)\tau_0 + N(\tau-\tau_0)),$ and we can again solve for $n_B(x)$ to find
\begin{align}
    \tilde n_{B,\pm} &= \frac{1}{2}\left(-1 + \frac{k+l}{\zeta} \pm \sqrt{\left(1 + \frac{k+l}{\zeta}\right)^2-4\frac{k}{\zeta}}\right)\\
    &= -\frac{1}{2} + \frac{k+l}{2\zeta} \pm \frac{1}{2\zeta}\sqrt{\left(\zeta + k+l\right)^2-4\zeta k}.
\end{align}
Here only the solution with positive occupation is valid $\tilde n_B = \tilde n_{B,-}\geq 0$.
Similarly to the Fermionic case, we can use solve for the temperature and obtain the MLE estimator.
The temperature estimate is then
\begin{align}
    {\tilde T}_{\rm LME}^{(b)} \coloneqq \frac{\epsilon}{\log \frac{1+{\tilde n}_B}{{\tilde n}_B}}.
\end{align}

\section{Comparison to equilibrium thermometry}\label{app:B}
The Fisher information of equilibrium thermometry, up to constants, for a 2-level energy spectrum reads~\cite{PhysRevLett.114.220405}
\begin{align}\label{eq:FI_eq_2lev}
    F^{\rm eq}_{2\rm lev}=\frac{N_0(N-N_0)}{\left(N_0 + (N-N_0)e^{-x}\right)^2}x^2 e^{-x},
\end{align}
where the dependence on $N$ and $N_0$ is different from the continuously monitored scenario (note the power two in the denominator). Taking $N_0=CN$, we can find that the best solution is to set $C(x) = (1+\exp(x))^{-1}$. By substitution we get
\begin{align}
    F^{\rm eq}_{2\rm lev} = \frac{C(x)(1-C(x))}{[C(x)+(1-C(x))e^{-x}]^2}x^2 e^{-x} N = \frac{x^2}{4}.
\end{align}
Note that this only scales with $x^2$. Apparently, this leads to an unbounded FI by setting $x\to\infty$. However, we are not allowed to do so, as this leads to $C(x)\to 0$, which means there is no ground state degeneracy, and the energy variance and the FI is zero for such a single level Hamiltonian structure. We have to demand that $1/N \leq C(x) < N$, such that we have two levels. Setting $C(x) = 1/N$ gives the solution $x=\log (N-1) \approx \log N$---any higher value for $x$ violates $1/N \leq C(x)$. Thus, the ultimate FI scales with $(\log N)^2$, and is achieved by a maximally degenerate excited state. See \cite{PhysRevLett.114.220405} for an alternative derivation.

Lastly, let us remark the difference between the ground state degeneracy in the equilibrium setting and the continuously monitoring setting. At thermal equilibrium, the FI \eqref{eq:FI_eq_2lev} is proportional to $p({\rm GS}) p(\rm ES)$---with $p({\rm GS}) = N_0 p_0^{eq}$ the probability of being in the ground state manifold, and  $p({\rm ES}) = (N-N_0) p_1^{eq}$ the probability of being in the exited state manifold. Hence this is optimal when $p({\rm GS})$ and $p({\rm ES})$ are both as close to $1/2$ as possible, which is obtained by the maximally degenerate setting. For continuous monitoring, it is slightly different. 
Since the dynamics is given by an order-$1$ Markov process with two energy levels, the Fisher information is given by Eq.~(15) in reference \cite{boeyens2023probe}. Extracting only the terms that depend on the degeneracy, this reads
\begin{align}
    {\tilde F}_{\rm 2lev} \propto \frac{N_0\Gamma_{10} (N-N_0)\Gamma_{01}}{N_0\Gamma_{10} +  (N-N_0)\Gamma_{01}} \propto \frac{N_0(N-N_0)}{N_0 + (N-N_0)e^{-x}},
\end{align}
where in the second proportionality we used detailed balance and got rid of terms that are independent of $N$ and $N_0$.
This matches our intuition that the FI is proportional to the likelihood of jumps, which depend on two things (i) what is the likelihood of being in either of the two manifolds $p({\rm GS})$ or  $p({\rm ES}) = (N-N_0) p_1^{eq}$. (ii) The degeneracy of the other manifold $N-N_0$ or $N_0$, respectively. Thus, ${\tilde F}_{\rm 2lev} \propto p({\rm GS})(N-N_0) \propto p({\rm ES})N_0 \propto N_0(N-N_0)/(N_0 + (N-N_0)e^{-x})$---which also matches Eqs. \eqref{eq:2level} and \eqref{eq:2levelboson}. Maximizing this quantity over $N_0$ gives a degenerate ground state, as we discussed in the main text.
\section{The empirical FI}\label{app:C}
According to \cite{smiga2023stochastic}, the empirical FI i.e., that of looking only at the average of the populations, for a system with detailed balance, reads
\begin{align}
    K = \frac{-\tau}{2} (\partial_Tp_{eq})^T P^{-1}W (\partial_Tp_{eq}),
\end{align}
where $P={\rm diag~}p_{\rm eq}$ is the diagonal matrix of the equilibrium populations. In our fermionic bath problem, for the 2-level energy structure, this simplifies to
\begin{align}
    {\tilde K}^{(f)}_{\rm 2lev} = \frac{ \,{N_0\left(N-N_0 \right)}}{\,{N_0+(N-N_0){\mathrm{e}}^{-x}}}\frac{x^2}{2\,{\left({\mathrm{e}}^x +1\right)}}.
\end{align}
The dependence on $N$ and $N_0$ are similar to the full statistics Fisher information rate. Therefore, by taking the $N_0 = CN$ ansatz, we have the optimal solution given by $C(x) = (\sqrt{\exp(x)} + 1)^{-1}$. Thus, we have
\begin{align}
    {\tilde K}_{2\rm lev}^{(f)} = \frac{C(x)(1-C(x))}{C(x) + (1-C(x))e^{-x}} \frac{x^2}{2\,{\left({\mathrm{e}}^x +1\right)}} N = f^{\prime}(x) N.
\end{align}
The maximum value of the function $f^{\prime}(x)$ can be found numerically. We find that
\begin{align}
    {\tilde K}_{2\rm lev}^{*(f)} & \approx 0.1448 N \nonumber\\
    x^* & \approx 2.7233\nonumber\\
    N_0^* & \approx \max\{{\rm floor}(0.2040 ~ N),~1\},
\end{align}

Likewise, in the bosonic case, for the 2-level ansatz we have
\begin{align}
    {\tilde K}^{(b)}_{2\rm lev} = \frac{N_0  \,{\left(N-N_0 \right)}}{{N_0 + (N-N_0)\,{\mathrm{e}}^{-x} }} \frac{\,x^{2+s}}{2\,{\left({\mathrm{e}}^x -1\right)}\,},
\end{align}
which by taking $N_0 = C N$, and optimising over $C$, which leads to $C(x) = (\sqrt{\exp(x)}+1)^{-1}$, gives
\begin{align}
    {\tilde K}_{2\rm lev}^{(b)} = \frac{C(x)(1-C(x))}{C(x) + (1-C(x))e^{-x}} \frac{x^{2+s}}{2\,{\left({\mathrm{e}}^x - 1\right)}} N = b^{\prime}(x) N.
\end{align}
We can summarise the optimal values in the table below
\begin{table}[h]
    \centering
    \begin{tabular}{cccc}
    \hline\hline
        SD & $x^{*}$ & $C^{*}$ & $b^{\prime}(x^*)$\\
        \hline\hline
        $s=1^+$ & 3.4079 & 0.1539 & 0.4851\\
        \hline
        $s=1.5$ & 3.9050 & 0.1243 & 0.9274\\
        \hline
        $s=2$ & 4.3850 & 0.1004 & 1.8879 \\
        \hline
        $s=3$ & 5.3215 & 0.0653 & 9.1514\\
        \hline
    \end{tabular}
    \caption{The optimal values of the gap and the corresponding ground state degeneracy fraction $C^*$ and the empirical FI coefficient $b^{\prime}(x^*)$. At the first row, $s=1^+$ is strictly bigger than one (the right limit).}
    \label{table:bos_opt_empirical}
\end{table}
\section{Robustness to energy structure}
\begin{figure}[h!]
    \centering
    \includegraphics[width=0.49\linewidth]{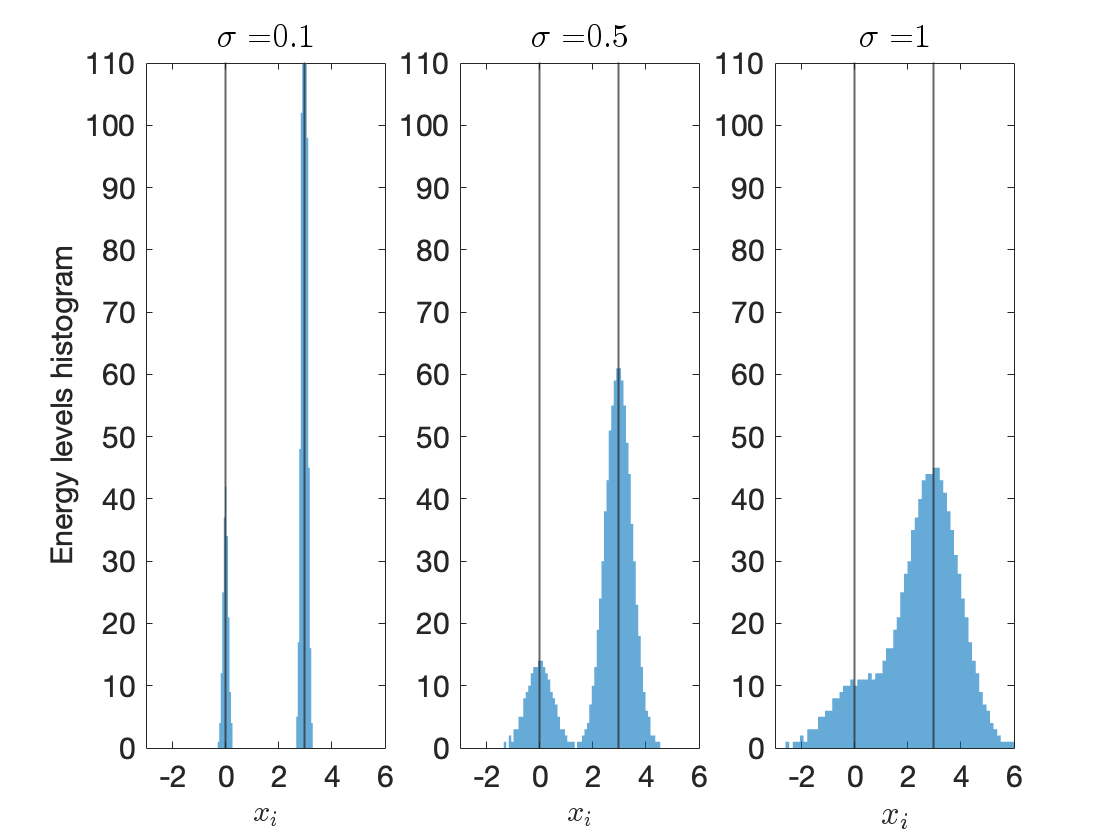}
    \includegraphics[width=0.49\linewidth]{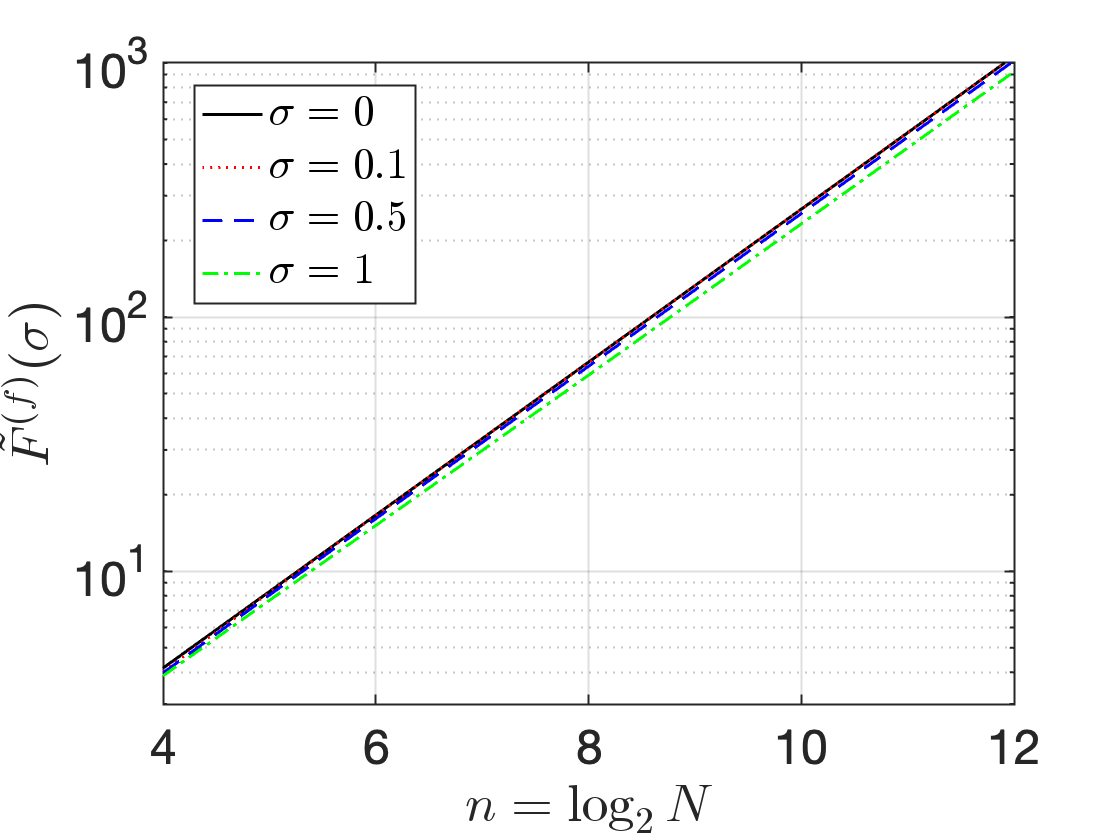}
    \caption{{\bf Left:} The energy structure chosen as two Gaussian distributions with standard deviation $\sigma$ and mean values that are the two optimal energies (i.e., $0$ and $x^*\approx 2.9682$). Here, we set $N=2^{10}$. {\bf Right:} The Fisher information of such Hamiltonian structures are quite close to the optimal two level setting, proving the robustness of our results.}
    \label{fig:robustness}
\end{figure}

In the main text we showed that by choosing the effective 2-level system we can obtain a Fisher information rate that scales linearly with $N$. We also argued that this scaling is robust against initial temperature ignorance, and thus can be also useful in a global strategy---as evidenced by the fact that the fisher information rate is proportional to $N$ for any choice of $x$. That means even if we did not choose the right gap, the scaling will survive. Here, we consider a secondary type of robustness. Particularly, we consider a scenario in which we cannot exactly prepare all the degenerate states at the same energy. To this aim, we take both the ground state manifold, and exited state manifold to be Gaussian distributed around the optimal values ($0$ and $x^*$, respectively), with a standard deviation $\sigma$. In what follows we mainly focus on the fermionic case, however, a similar behavior for the bosonic case is expected. We depict this in Fig.~\ref{fig:robustness}. Evidently, even for large values of $\sigma$, the scaling with $n$ is preserved to a good extent, hence assuring the robustness of the optimal 2-level setting against energy fluctuations within the two manifolds.

On a more analytical level, it can also be argued, why a distribution of the energy levels that differs from the optimal effective 2-level structure can be sufficient for attaining a scaling close to the optimal one.
In the optimal case from Eq.~\eqref{eq:2level} obtained with the 2-level ansatz, there are $N_0$ ground state levels and $N-N_0$ excited state levels with the single dimensionless gap $x^*$ that we determined numerically.
Furthermore, in the optimal case, the distribution of excited versus ground state levels is given by $N^*_0=N/(1+e^{x^*/2})$.
In a suboptimal case, however, we would expect that the energy levels spread around their optimal values.
To be explicit, we assume that at least a fraction of $qN$ states with $0<q$ is within a symmetric window of width $\sigma$ around the optimal ground state energy,
and at least a fraction of $pN$ states with $0<p,$ and $q+p\leq 1$, is within a symmetric window of width $\sigma$ around the optimal excited state energy $x^*$.
The width $\sigma$ is an energy normalized by the inverse temperature like the parameter $x^*$.
The remaining states may have energy higher than the $x^*+\sigma$, but it is important that none have energy smaller than $-\sigma/2$ to avoid the system to `freeze' in a very low ground state.
Under these assumptions, the asymptotic Fisher information rate scaling with $N$ can be recovered, but at the loss of constant factors. Here, we focus on the derivation for the fermionic case, though the bosonic case follows analogously, but with different constant factors.
We start by recalling the general expression~\eqref{eq:FIrate} for the fermionic Fisher information rate, which can be written as,
\begin{align}
    {\tilde F}^{(f)} &=  \sum_i p^{eq}_i \sum_{j\neq i}x_{ji}^2\frac{e^{2 x_{ji}}}{[1+e^{x_{ji}}]^3}.
\end{align}
Since all terms in the sum for the Fisher information rate are positive, it can be lower bounded by the terms that lie within a window of width $\sigma$ within the optimal energies, while all others are dropped.
Formally, we define by $\mathcal E_{\rm low} = \{i : x_i \in [-\sigma/2,\sigma/2]\}$ the set of states that are $\sigma$-close to the optimal lower energy, and $\mathcal E_{\rm high}= \{i : x_i \in [x^*-\sigma/2,x^*+\sigma/2]\}$ the set of states $\sigma$-close to the upper energy.
That also implies that $|\mathcal E_{\rm low}|\geq q N$ and $|\mathcal E_{\rm high}| \geq pN$, where $|\mathcal E|$ denotes the number of elements of the set $\mathcal E$.
As long as the lower energies and the upper energies have no overlap $\mathcal E_{\rm low} \cap \mathcal E_{\rm high}=\emptyset$, which is guaranteed if $\sigma < x^*,$ a lower bound for ${\tilde F}^{(f)}$ is given by,
\begin{align}
\label{eq:tildeF_lowerbound1}
    {\tilde F}^{(f)} &\geq \sum_{i \in \mathcal E_{\rm low},\,j\in \mathcal E_{\rm high}} \left( p_i^{eq} x_{ji}^2\frac{e^{2 x_{ji}}}{[1+e^{x_{ji}}]^3}+p_j^{eq} x_{ji}^2 \frac{e^{-2 x_{ji}}}{[1+e^{-x_{ji}}]^3}\right).
\end{align}
Every term in the sum can now be explicitly bounded. Firstly, starting with the equilibrium distribution in the low energy manifold,
\begin{align}
    \min_{i\in\mathcal E_{\rm low}} p_i^{eq} &\geq \frac{e^{-\sigma/2}}{(1-p)N e^{\sigma/2} + p Ne^{-x^* + \sigma/2}} =\frac{ce^{-\sigma}}{N},
\end{align}
where $c=1/(1-p+pe^{-x^*})>1$ is a constant independent of $N$ and given by a function of $x^*$ and the fraction $p$.
Similarly for the excited state manifold,
\begin{align}
    \quad \min_{i\in\mathcal E_{\rm high}} p_i^{eq} &\geq \frac{c e^{-x^*-\sigma}}{N}, 
\end{align}
with $c>1$ the same constant independent of $N$ as before.
Also the terms $x_{ji}^2 e^{\pm 2x_{ji}}/(1+e^{\pm x_{ji}})^3$ in the sum can be lower bounded by a non-zero constant,
\begin{align}
    c_{\pm}\coloneqq \min_{x\in [x^*-\sigma,x^* +\sigma]} x^2 \frac{e^{\pm 2 x}}{[1+e^{\pm x}]^3} >0,
\end{align}
so long as $x^*>\sigma$ which coincides with the assumption previously made that $\mathcal E_{\rm low} \cap \mathcal E_{\rm high}=\emptyset$.
As a result, we can further bound the expression in Eq.~\eqref{eq:tildeF_lowerbound1}, showing that the Fisher information rate asymptotically still scales with $N$, albeit with worse constant factors than in the case with a true 2-level structure,
\begin{align}
    {\tilde F}^{(f)} &\geq qpN^2  \frac{c e^{-\sigma}(c_+ + c_-e^{-x^*})}{N} \\
    &\geq \Omega(N).
\end{align}
Here, $\Omega$ is the Knuth-notation for an asymptotic lower bound up to constant factors~\cite{Knuth1976}.
This shows that in the Fermionic case the Fisher information rate still scales asymptotically linearly in $N$, so long as $\sigma$ is sufficiently smaller than the energy gap $x^*$ and the fraction of states within that window is constant and independent of $N$.
The argument also shows the limits where the asymptotic scaling can not be upheld anymore.
For example, when the fraction of states within a small enough window $\sigma$ decreases as a function of $N$ because of, e.g., energy level repulsion, the optimal scaling can break down.
So long as a constant fraction stays within the desired windows, and no state lies below the lower energy window, however, the results are robust.

\end{document}